\documentclass[conference]{IEEEtran}
\IEEEoverridecommandlockouts
\usepackage{cite}
\usepackage{amsmath,amssymb,amsfonts}
\usepackage{algorithmic}
\usepackage[ruled,linesnumbered]{algorithm2e}
\usepackage{amsthm}
\usepackage{graphicx}
\usepackage{textcomp}
\usepackage{multirow}
\usepackage{makecell}
\usepackage{xcolor}
\def\BibTeX{{\rm B\kern-.05em{\sc i\kern-.025em b}\kern-.08em
    T\kern-.1667em\lower.7ex\hbox{E}\kern-.125emX}}
\begin{document}

\title{Application of Data-driven Model Predictive Control for Autonomous Vehicle Steering\\
\thanks{This work was supported in part by the Shanghai Scientific Innovation Foundation (No.23DZ1203400), the Belt and Road Cooperation Program under the 2023 Shanghai Action Plan for Science, Technology and Imnovation (23210750500), the National Natural Science Foundation of China (52302502), the State Key Laboratory of Intelligent Green Vehicle and Mobility under Project No. KFZ2408, the Young Elite Scientists Sponsorship Program by CAST (2022QNRC001), and the Fundamental Research Funds for the Central Universities.

}
\thanks{*Corresponding author.
}
}

\author{\IEEEauthorblockN{1\textsuperscript{st} Jiarui Zhang}
\IEEEauthorblockA{\textit{Department of Traffic Engineering \&} \\
\textit{Key Laboratory of} \\
\textit{Road and Traffic Engineering,} \\
\textit{Ministry of Education} \\
\textit{Tongji University}\\
Shanghai, China \\
zjr0915@tongji.edu.cn}

\and
\IEEEauthorblockN{2\textsuperscript{nd} Aijing Kong}
\IEEEauthorblockA{\textit{Department of Traffic Engineering \&} \\
\textit{Key Laboratory of} \\
\textit{Road and Traffic Engineering,} \\
\textit{Ministry of Education} \\
\textit{Tongji University}\\
Shanghai, China \\
ajkong@tongji.edu.cn}

\and
\IEEEauthorblockN{3\textsuperscript{rd} Yu Tang}
\IEEEauthorblockA{\textit{China Automotive Engineering} \\
\textit{Research Institute Co., Ltd.,} \\
Chongqing, China \\	
tangyu@caeri.com.cn}

\and
\IEEEauthorblockN{4\textsuperscript{th} Zhichao Lv}
\IEEEauthorblockA{\textit{Department of Traffic Engineering \&} \\
\textit{Key Laboratory of} \\
\textit{Road and Traffic Engineering,} \\
\textit{Ministry of Education} \\
\textit{Tongji University}\\
Shanghai, China \\
lvzhichao@tongji.edu.cn}

\and
\IEEEauthorblockN{5\textsuperscript{th} Lulu Guo}
\IEEEauthorblockA{
\textit{Department of Control Science} \\
\textit{and Engineering} \\
\textit{Tongji University}\\
Shanghai, China \\
guoll21@tongji.edu.cn}

\and
\IEEEauthorblockN{6\textsuperscript{th} Peng Hang*}
\IEEEauthorblockA{\textit{Department of Traffic Engineering \&} \\
\textit{Key Laboratory of} \\
\textit{Road and Traffic Engineering,} \\
\textit{Ministry of Education} \\
\textit{Tongji University}\\
Shanghai, China \\
hangpeng@tongji.edu.cn}
}
\maketitle

\begin{abstract}
With the development of autonomous driving technology, there are increasing demands for vehicle control, and MPC has become a widely researched topic in both industry and academia. Existing MPC control methods based on vehicle kinematics or dynamics have challenges such as difficult modeling, numerous parameters, strong nonlinearity, and high computational cost. To address these issues, this paper adapts an existing Data-driven MPC control method and applies it to autonomous vehicle steering control. This method avoids the need for complex vehicle system modeling and achieves trajectory tracking with relatively low computational time and small errors. We validate the control effectiveness of the algorithm in specific scenario through CarSim-Simulink simulation and perform comparative analysis with PID and vehicle kinematics MPC, confirming the feasibility and superiority of it for vehicle steering control.
\end{abstract}


\section{Introduction}
Currently, autonomous driving has matured and is gradually coming into the public eye. Numerous internet and vehicle manufacturing companies are investing increasing efforts into researching autonomous driving technology, which has significantly contributed to improving traffic congestion, reducing traffic accidents, and enhancing economic benefits \cite{al2021impacts, garg2023can}. Mastinu et al. analyzed the reasons and scenarios in which drivers lose control of the vehicle. They pointed out that after severe lane changes, gusts of wind, or other disturbances, drivers might be unable to regain the intended actions, potentially posing traffic safety hazards \cite{mastinu2024drivers}. Moreover, Ahangar et al. found that the number of fatalities due to road traffic accidents is continually rising, with many accidents resulting from driver fatigue and distraction \cite{ahangar2021survey}. Additionally, the proportion of carbon dioxide emissions from road traffic in the total human carbon dioxide emissions is also increasing \cite{li2021forecast}. Therefore, research on autonomous driving technology is urgently needed.

Autonomous vehicles are composed of multiple modules, including perception, prediction, planning, decision-making, and control. Among them, control is one of the most critical modules, and the control methods have always been a key research focus. In the field of autonomous driving control, PID and adaptive control, etc. are widely used in the industry, which do not require mathematical modeling of the system, and the optimal control quantity can be obtained using the vehicle's state and reference trajectory \cite{li2024time, petrov2014modeling}. In academia, however, Model Predictive Control (MPC) is a widely researched topic. First proposed by Richalet et al. in 1978 \cite{richalet1978model}, MPC have since evolved with various modifications to suit different control scenarios and controlled objects \cite{donkers2012stability, hang2021path, hang2021active}.

However, general MPC require precise modeling of the controlled system. Currently, most MPC methods for autonomous driving steering control are based on vehicle kinematics or dynamics models \cite{tang2020improved, lai2021comparative}. Vehicle kinematics models have fewer parameters and simple structures, but they simplify the autonomous vehicle into a two-wheel model, which significantly deviates from the actual vehicle situation and results in low control precision. On the other hand, vehicle dynamics models contain more parameters and require parameter calibration through experiments, and also the strong nonlinearity of the models leads to high optimization computational costs. Therefore, researchers have begun considering how to avoid the cumbersome modeling process and directly use data for system characteristics analysis \cite{jian2009notes} and controller design \cite{van2020data, de2019formulas, breschi2023data}.

Currently, Data-driven MPC becomes widely researched. This control method avoids the precise modeling of system, as required by traditional MPC algorithms, and reduces computational time while maintaining high control accuracy. To address existing vehicle control problems, this paper studies the existing Data-driven MPC and, by integrating vehicle system characteristics, applies it to vehicle steering control and verifies the feasibility of this method. The contributions of this paper are presented as follows:

1. Based on the research of \cite{coulson2019data, berberich2020trajectory, berberich2020robust, berberich2020data}, a Data-driven Model Predictive Control method is applied to autonomous vehicle steering control.

2. The feasibility of the application of DDMPC to autonomous vehicle steering was verified through simulation experiments, and the superiority of this algorithm was demonstrated by comparing it with other algorithms.

The rest of the paper is organized as follows. Section II provides a brief introduction to our research problem and discusses Willems' Lemma. In Section III, we introduce the existing research on DDMPC. Based on this, we make minor modifications to make the algorithm applicable to autonomous vehicle  control. In Section IV, we validate the effectiveness of the proposed algorithm through CarSim and Simulink simulation experiments and conduct a comparative analysis with PID and vehicle kinematics MPC algorithms. Finally, conclusions are drawn in Section V.

\begin{figure}[htbp]
  \begin{center}
  \centerline{\includegraphics[width=3.5in]{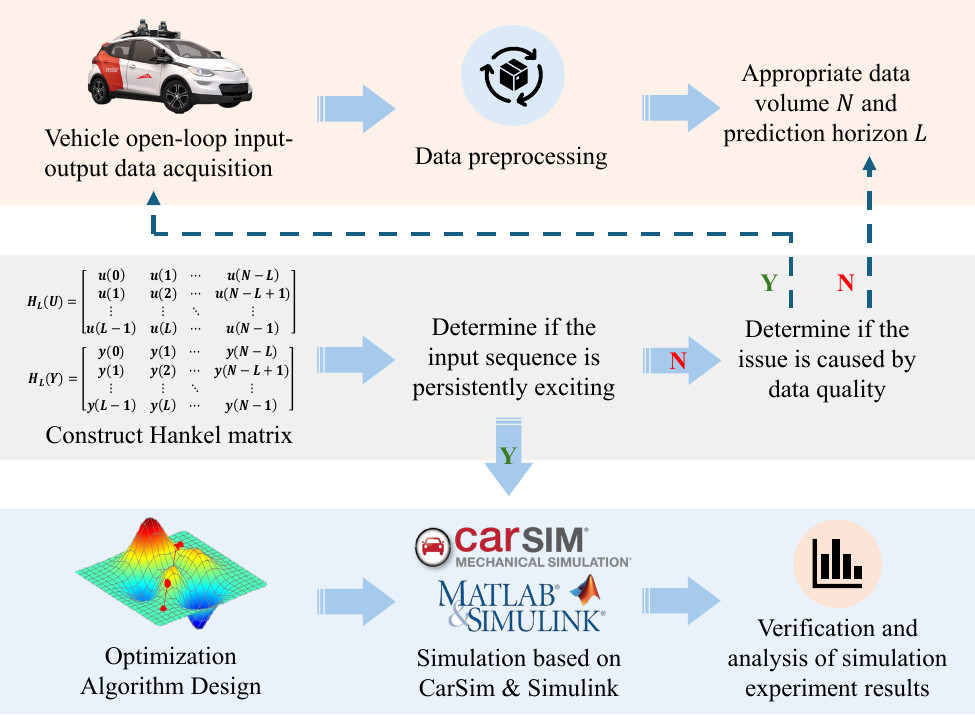}}
  \caption{Algorithm application process of the DDMPC for AV steering control.}
  \label{fig 1}
  \end{center}
\end{figure}

\section{Problem Statement}

The development of autonomous vehicle technology relies on efficient and reliable control algorithms. The advantage of MPC lies in its ability to calculate high-precision control inputs within a limited prediction horizon, based on the vehicle model and reference trajectory. Consequently, MPC often depends on accurate vehicle models, but modeling and parameter calibration of traditional vehicle models—especially dynamic models—become extremely challenging. Additionally, vehicle models often have many parameters and strong nonlinearity, which may consume a significant amount of computational time during optimization. Most scholars and engineers address this issue by linearization, but this often leads to a decrease in model accuracy.

Based on the Willems's lemma, which is a data-based method for system identification \cite{willems2005note}, Jeremy first proposed an algorithmic framework called Data-enabled Predictive Control and applied it on aerial robotics\cite{coulson2019data}. Thereafter, Berberich et al. designed a robust Data-driven MPC control method \cite{berberich2020trajectory, berberich2020robust, berberich2020data}. This method can directly use the Hankel matrix constructed from offline input-output trajectory data of the system to replace complex system models, predicting future states of the system and thereby calculating the optimal control inputs. Lu et al. used this method to complete the data-driven identification of vehicle and designed a DDMPC controller for vehicle lateral stability control\cite{lu2023design}. Subsequently, many scholars have expanded and applied this method \cite{schmitz2022willems, torrente2021data}.

We build on this foundation by applying the data-driven MPC algorithm proposed by \cite{coulson2019data} and \cite{berberich2020trajectory} to steering control of autonomous vehicles and provide the algorithm application flowchart, as shown in Fig.~\ref{fig 1}.

\section{Application of DDMPC for Autonomous Vehicle Steering}

\subsection{Willems' Lemma and Application}\label{AA}
Here, we first review the description and application of Willems' Lemma by \cite{coulson2019data} and \cite{berberich2020trajectory}.

Suppose the dynamic behavior of a system is described by the following input-output relationship expressed by Eq.~\ref{eq 1}.

\begin{equation}
\begin{split}
    y(t) = G(u(t))
    \label{eq 1}
\end{split}
\end{equation}
where $u(t) \in U \subset \mathbb{R}^m$ is the system input at time $t$, with $m$ being the input dimension; $y(t) \in Y \subset \mathbb{R}^p$ is the system output at time $t$, with $p$ being the output dimension; $G(\cdot)$ is the system behavior model, generally represented by a transfer function or state-space equations.

Apply a set of inputs $U$ to the system, which correspondingly generates a set of outputs $Y$. The collected open-loop input-output data are represented as two sets of vectors in Eq.~\ref{eq 2}.

\begin{equation}
\begin{split}
    U &= [u(0), u(1), \cdots, u(N-1)] \\
    Y &= [y(0), y(1), \cdots, y(N-1)]
    \label{eq 2}
\end{split}
\end{equation}
where $N$ is the number of data sets, and the selection of this parameter will directly influence the subsequent design of the Data-driven MPC.

Process the collected input-output data, which mainly includes data cleansing, continuity checking and noise removal, etc. Then, extend the data into Hankel matrices. The resulting order $L$ Hankel matrix is as Eq.~\ref{eq 3} and Eq.~\ref{eq 4}.

\begin{equation}
\begin{split}
    H_L(U) = \begin{bmatrix}
    u(0) & u(1) & \cdots & u(N-L) \\
    u(1) & u(2) & \cdots & u(N-L+1) \\
    \vdots & \vdots & \ddots & \vdots \\
    u(L-1) & u(L) & \cdots & u(N-1)
    \end{bmatrix}
    \label{eq 3}
\end{split}
\end{equation}

\begin{equation}
\begin{split}
    H_L(Y) = \begin{bmatrix}
    y(0) & y(1) & \cdots & y(N-L) \\
    y(1) & y(2) & \cdots & y(N-L+1) \\
    \vdots & \vdots & \ddots & \vdots \\
    y(L-1) & y(L) & \cdots & y(N-1)
    \end{bmatrix}
    \label{eq 4}
\end{split}
\end{equation}
where $L$ is the basic prediction horizon of the MPC algorithm. For the input matrix $H_L (U)$, we determine whether the input sequence $U$ satisfies the requirement of persistent excitation based on the following definition.

\textbf{Definition in \cite{coulson2019data, berberich2020trajectory}}: The input sequence $U$ is order $L$ persistently exciting if and only if the rank of the order $L$ Hankel matrix $H_L (U)$ constructed from this input sequence satisfies the Eq.~\ref{eq 5}.

\begin{equation}
\begin{split}
    \text{rank}[H_L(U)] = mL
    \label{eq 5}
\end{split}
\end{equation}

This means that the input sequence is rich enough to excite all dynamic modes of the system. Based on this definition, we can further explore the relationship between the input-output Hankel matrix and the system under study for predicting system outputs.

\textbf{Willems' Lemma\cite{willems2005note}}: If $\{U,Y\} = \{u_k, y_k\}_{k=0}^{N-1}$ is a set of $N$ input-output data measured from system $G$, and $U$ is order $L+n$ persistently exciting, then $\{\bar{u}_k, \bar{y}_k\}_{k=0}^{L-1}$ are the predicted input-output sequences of system $G$ for the future $L$ time steps based on $\{U,Y\}$, if and only if there exists an $\alpha \in \mathbb{R}^{N-L+1}$ that satisfies Eq.~\ref{eq 6}.

\begin{equation}
\begin{split}
    \begin{bmatrix}
    H_L(U) \\
    H_L(Y)
    \end{bmatrix} \alpha = \begin{bmatrix}
    \bar{u} \\
    \bar{y}
    \end{bmatrix}
    \label{eq 6}
\end{split}
\end{equation}
where $n$ is the number of states of the controlled system. According to the previously given definition of persistent excitation, the rank of the order $L+n$ Hankel matrix $H_{L+n} (U)$ constructed from the input sequence $U$ of length $N$ must satisfy the Eq.~\ref{eq 7}.

\begin{equation}
\begin{split}
    \text{rank}[H_{L+n}(U)] = m(L+n)
    \label{eq 7}
\end{split}
\end{equation}

Therefore, with known historical input-output data, if an optimal $\alpha$ can be found, we can use the above lemma to predict the system's future inputs and outputs. This lemma is of great interest in the fields of system identification and data-driven control because it provides a new approach that allows us to bypass the derivation of the system model (and even the knowledge of the specific form of the system) and directly obtain dynamic behavior information from the open-loop data generated by the system\cite{coulson2019data, berberich2020data, berberich2020robust, berberich2020trajectory}. This is because, when the input signal meets certain persistent excitation conditions, the Hankel matrix constructed from a pre-collected known input-output data segment implicitly represents the system's dynamic characteristics and can be used to represent any input-output trajectory of the system of the basic prediction horizon $L$ through linear combinations.

\subsection{DDMPC for Vehicle Steering Control}\label{BB}
Based on Willems' lemma, existing research combined it with MPC roll optimization and designed the DDMPC optimization model\cite{coulson2019data, berberich2020trajectory}, and we adapt it in this subsection for vehicle steering control as shown in Eq.~\ref{eq 8} to Eq.~\ref{eq 12}.

\begin{equation}
\begin{split}
    J_L(\bar{u}(t), \bar{y}(t)) = \sum_{k=0}^{L-1}  \| \bar{y}_k(t) - y_k^r(t) \|_Q^2  \\
    + \| \bar{u}_k(t) - u_k^r(t) \|_R^2 + \lambda \| \alpha(t) \|_2^2
    \label{eq 8}
\end{split}
\end{equation}
where $y_k^r(t)$ and $u_k^r(t)$ represent the expected output and input of the system at time $t$ for $k$ time steps ahead from the current state; $Q \in \mathbb{R}^{p \times p}$ and $R \in \mathbb{R}^{m \times m}$ are symmetric positive definite matrices. In the context of autonomous driving control, $y_k^r(t)$ corresponds to the trajectory point information of the vehicle's position and state at the current time $t$ extended $k$ time steps ahead; $u_k^r(t)$, from the perspective of safety and comfort, is generally set to be a zero vector, indicating that the desired control action should be minimized as much as possible. Based on Willems' lemma, we describe $\bar{y}$ and $\bar{u}$ using the Hankel matrix and constrain the system state, thereby forming the equality and inequality constraint equations for the following optimization problem.

\begin{equation}
\begin{split}
    \begin{bmatrix}
    H_n(U) \\
    H_n(Y)
    \end{bmatrix} \alpha(t) =
    \begin{bmatrix}
    u_{-n}(t) \\
    y_{-n}(t)
    \end{bmatrix}
    \label{eq 9}
\end{split}
\end{equation}

\begin{equation}
\begin{split}
   \begin{bmatrix}
    H_L(U) \\
    H_L(Y)
    \end{bmatrix} \alpha(t) =
    \begin{bmatrix}
    \bar{u}(t) \\
    \bar{y}(t)
    \end{bmatrix}
    \label{eq 10}
\end{split}
\end{equation}

\begin{equation}
\begin{split}
   \begin{bmatrix}
    \alpha_{\min} \\ 
    u_{\min} \\ 
    y_{\min} 
    \end{bmatrix}
    \leq
    \begin{bmatrix}
    \alpha(t) \\ 
    \bar{u}(t) \\ 
    \bar{y}(t) 
    \end{bmatrix}
    \leq
    \begin{bmatrix}
    \alpha_{\max} \\ 
    u_{\max} \\ 
    y_{\max} 
    \end{bmatrix}
    \label{eq 11}
\end{split}
\end{equation}

\begin{equation}
\begin{split}
    \bar{u}_k(t) \in U, \quad \bar{y}_k(t) \in Y, \quad k \in \mathbb{I}_{[0,L-1]}
    \label{eq 12}
\end{split}
\end{equation}

In this context, Eq.~\ref{eq 9} represents the initial constraint, where $u_{-n}(t)$ and $y_{-n}(t)$ denote the actual control inputs and outputs of the system for $n$ time steps before time $t$, respectively. This constraint aims to use the actual dynamic behavior of the system to constrain the decision variable $\alpha(t)$, ensuring that the prediction results closely match the actual system behavior. 

Eq.~\ref{eq 10} is the application of Willems' lemma in the Data-driven MPC algorithm. It uses the Hankel matrix to predict the system's input-output, serving as the equality constraint of this optimization problem. By comparing the predicted system output $\bar{y}(t)$ with the reference trajectory $y_k^r(t)$, the optimal control sequence $\bar{u}(t)$ is obtained.

Eq.~\ref{eq 11} represents the upper and lower bounds constraints on the decision variables. Unlike general systems, in the context of autonomous driving control, constraints on the control input $\bar{u}(t)$ are particularly critical because the control inputs are typically acceleration or front wheel steering angles. We generally need to limit the range of $\bar{u}(t)$ from the perspectives of both safety and comfort. It is worth noting that the prediction horizon chosen here should be $L+n$, which requires the open-loop input sequence to be order $L+2n$ persistently exciting\cite{berberich2020trajectory}. Additionally, the Hankel matrix constructed from the open-loop data should be of order $L+n$, satisfying Eq.~\ref{eq 13}.

\begin{equation}
\begin{split}
    H_{L+n}(\cdot) =
    \begin{bmatrix}
    H_n(\cdot) \\
    H_L(\cdot)
    \end{bmatrix}
    \label{eq 13}
\end{split}
\end{equation}

At this point, we have obtained the DDMPC framework for vehicle steering control. We only need to solve the optimization problem Eq.~\ref{eq 8} to Eq.~\ref{eq 12} at time $t$, and apply the first control input of the optimal control sequence $\bar{u}_1(t)$ (In this study, it is the left and right front wheel angle values) to the autonomous vehicle to complete the control at the current time.

\section{Experiments and Results}

To verify the effectiveness of the algorithm, simulation experiments are conducted based on CarSim and Simulink softwares. CarSim is a commonly used simulation platform for vehicle experiments, providing precise vehicle dynamics models which can realistically reproduce vehicle operations in real-world scenarios. The simulation experiments are divided into two stages: open-loop data collection and closed-loop algorithm simulation verification, following the same logic as the design of algorithm.

In CarSim, a D-Class Sedan vehicle model is selected as the test vehicle for open-loop data collection and algorithm simulation verification. The specific parameters are shown in Table.~\ref{table 1}.

\begin{table}[htbp]
\caption{Basic Parameters of the Vehicle Used for Simulation Tests}
\begin{center}
\begin{tabular}{|c|c|c|}
\hline
\textbf{Basic Vehicle Parameters} & \textbf{Parameter Value} & \textbf{Unit} \\
\hline
Sprung mass & 1370 & kg \\
\hline
Wheelbase & 2910 & mm \\
\hline
Height for animator & 1380 & mm \\
\hline
Width for animator & 2162 & mm \\
\hline
Yaw inertia & 2315.3 & kg$\cdot$m\(^2\) \\
\hline
\end{tabular}
\label{table 1}
\end{center}
\end{table}

\subsection{Open-Loop Data Collection}

In CarSim, we set up a test environment for open-loop data collection. This environment should include as many steering scenarios as possible to ensure that the input sequence is sufficiently rich to activate the dynamic characteristics of the vehicle. 

The vehicle's horizontal coordinate $X$ and vertical coordinate $Y$ are chosen in the global coordinate system, as well as the vehicle's heading angle $\phi$, as the output variables. The left front wheel steering angle $\delta_L$ and the right front wheel steering angle $\delta_R$ are selected as the control variables. Thus, the vehicle's open-loop input-output sequence can be expressed as Eq.~\ref{eq 14} and Eq.~\ref{eq 15}.

\begin{equation}
\begin{split}
    u(k) = \begin{bmatrix}
    \delta_L(k) \\
    \delta_R(k)
    \end{bmatrix}, \quad k = 0, 1, \cdots, N-1
    \label{eq 14}
\end{split}
\end{equation}

\begin{equation}
\begin{split}
    y(k) = \begin{bmatrix}
    x(k) \\
    y(k) \\
    \phi(k)
    \end{bmatrix}, \quad k = 0, 1, \cdots, N-1
    \label{eq 15}
\end{split}
\end{equation}

Due to the non-constant speed of the vehicle, it is necessary to interpolate the sparser parts of the collected raw data. Additionally, the data needs to be preprocessed, including data cleaning and outlier removal. The final amount of data obtained is $N = 646$.

\subsection{Simulation Experiments}

Since we do not need to model the vehicle system, the system order $n$ is unknown here. Therefore, we assign an upper bound $v$ to the system order, and set $v=6$ to substitute for $n$ in the algorithm. Additionally, we set the basic prediction horizon $L=24$, making the prediction horizon $L+v=30$. The weight matrices are set as $Q = I_p$, $R = 10^{-2} \cdot I_m$, and $\lambda = 1 \cdot 10^{-3}$. To meet the vehicle's safety and comfort requirements, we set $u_{\min}=-1.5^\circ$ and $u_{\max}=+1.5^\circ$. Besides, We don't bind $\alpha(t)$ and $\bar{y}(t)$, and keep the vehicle speed around $36 \,\text{km/h}$.

\begin{figure}[htbp]
  \begin{center}
  \centerline{\includegraphics[width=3.5in]{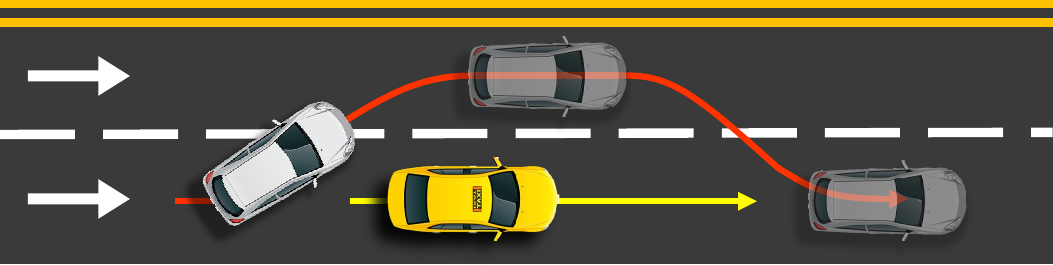}}
  \caption{Schematic diagram of a dual-lane switching scenario.}
  \label{fig 2}
  \end{center}
\end{figure}

A dual-lane switching scenario shown in Fig.~\ref{fig 2} is selected as the simulation experiment case. Additionally, the experiments are conducted using PID and vehicle kinematics MPC control algorithms in the same scenario and the comparative analysis is performed with the results of the Data-driven MPC experiments to demonstrate the advantages of the application of it in vehicle steering control.

\begin{figure*}[htbp]
  \begin{center}
  \centerline{\includegraphics[width=7in]{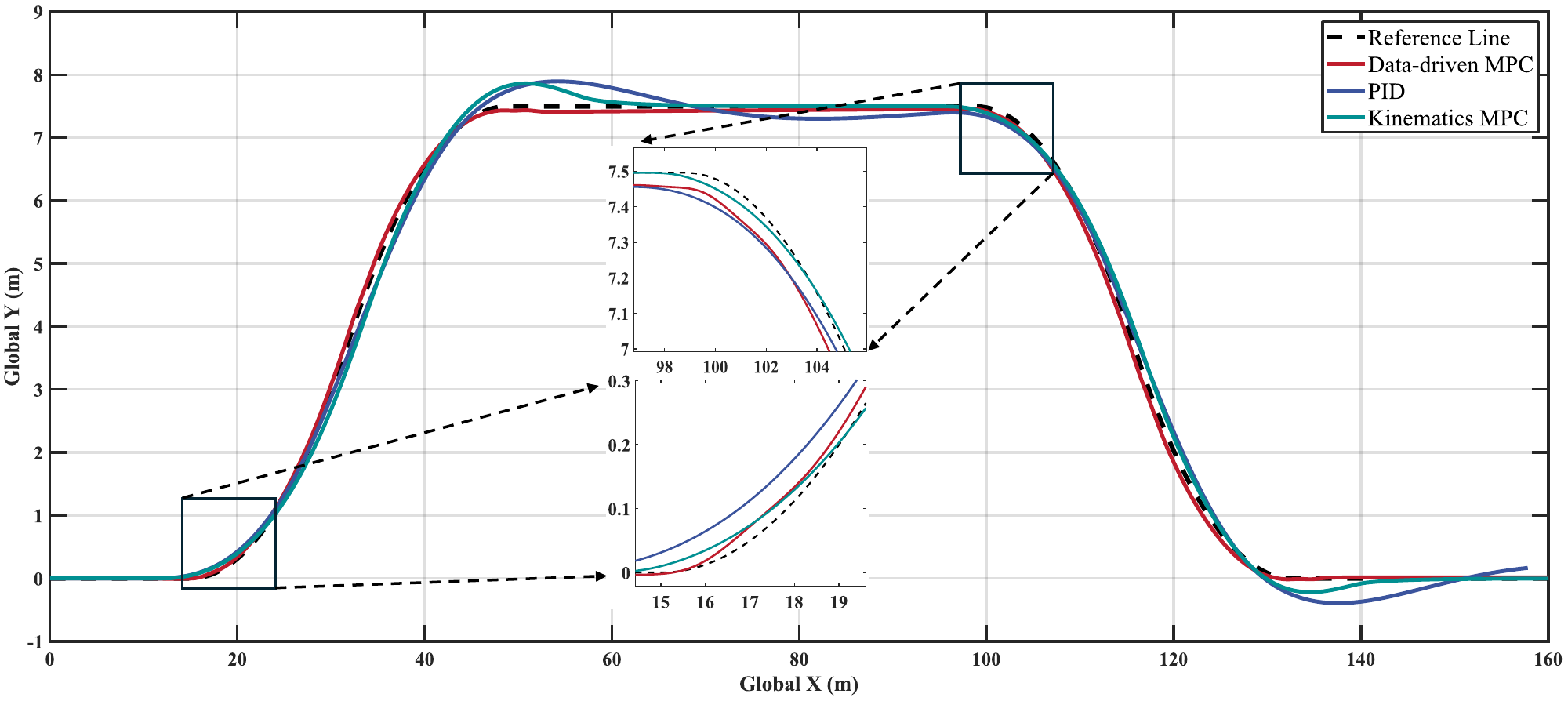}}
  \caption{Trajectories of Data-driven MPC, PID and kinematics MPC with reference line.}
  \label{fig 3}
  \end{center}
\end{figure*}

\subsection{Results Analysis}

\begin{figure}[htbp]
  \begin{center}
  \centerline{\includegraphics[width=3.5in]{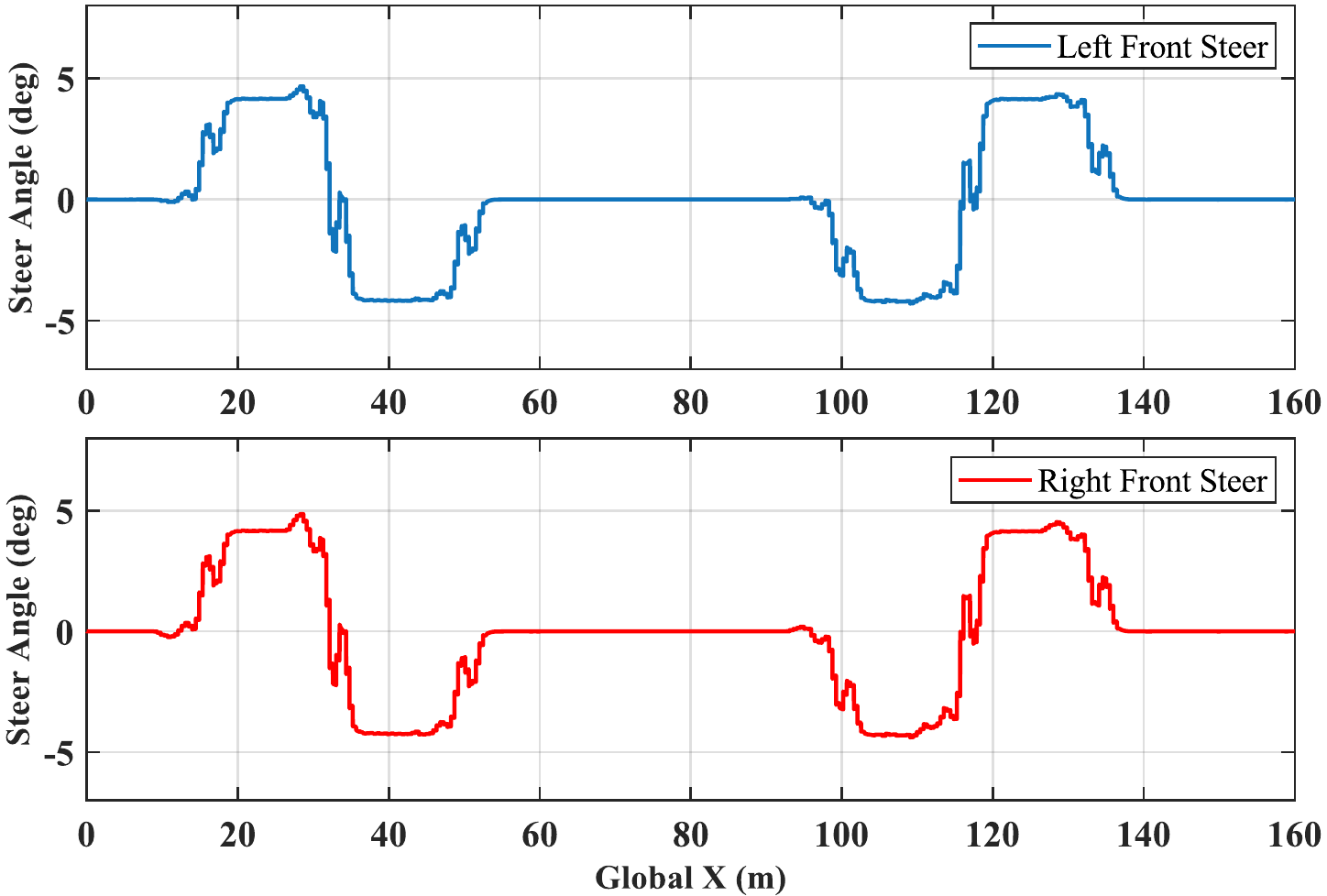}}
  \caption{Front wheel steering.}
  \label{fig 4}
  \end{center}
\end{figure}

Fig.~\ref{fig 4} shows the variation of the left and right front wheel steering angles over time under the control of DDMPC. As can be seen, the trends of the left and right front wheel steering angles are almost identical and change smoothly without severe fluctuations, remaining $-5^\circ$ to $+5^\circ$. This indicates that the algorithm allows for stable lane changes without excessively aggressive steering maneuvers, ensuring vehicle stability and comfort. Moreover, DDMPC can respond quickly in turning scenarios, ensuring vehicle control safety during emergency lane changes.

\begin{figure}[htbp]
  \begin{center}
  \centerline{\includegraphics[width=3.465in]{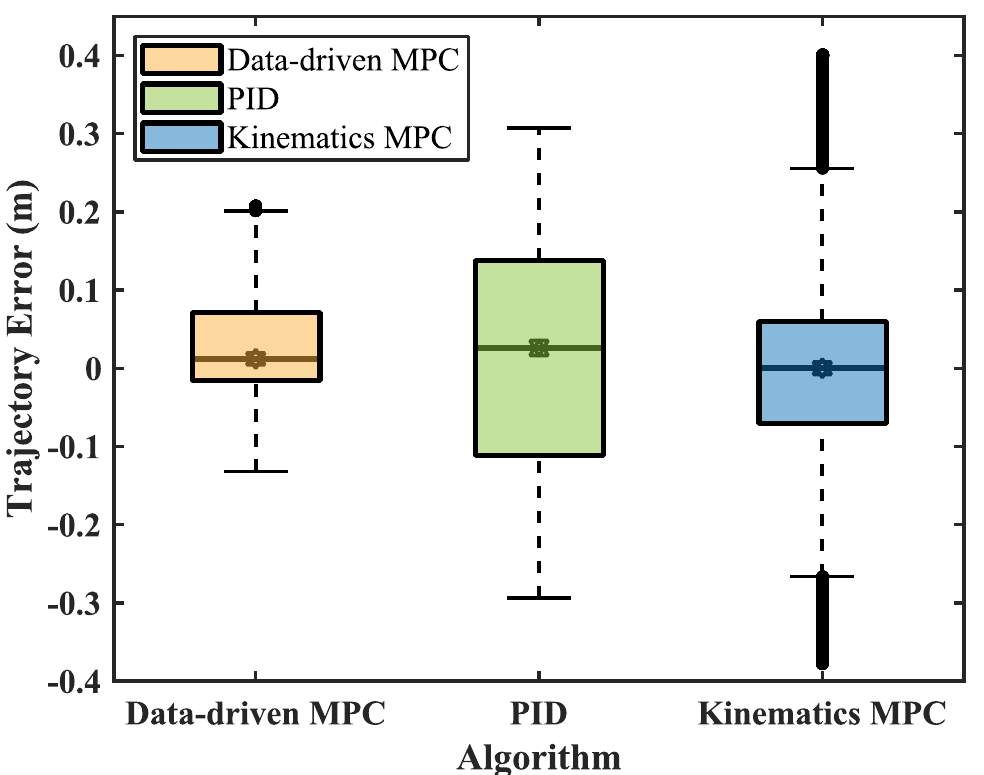}}
  \caption{Tracking error of Data-driven MPC, PID and kinematics MPC.}
  \label{fig 5}
  \end{center}
\end{figure}

To verify the practical value for steering control of DDMPC, we compared it with PID and vehicle kinematics MPC algorithms. Fig.~\ref{fig 3} and Fig.~\ref{fig 5} show the global trajectory and tracking error of the three algorithms in the same scenario. All three algorithms can achieve trajectory tracking within a certain error range. However, the vehicle kinematics MPC and PID algorithms exhibit vehicle deviation from the track during the curve-to-straight transition, whereas DDMPC ensures a quick response in steering angles during this transition, allowing the vehicle to follow the desired trajectory continuously. Additionally, as shown in Fig.~\ref{fig 5}, the error variation range of DDMPC is relatively small and generally maintained between $-0.1\, \text{m}$ and $0.2\, \text{m}$ with almost no outliers, which indicates that there are almost no severe deviations.

\begin{figure}[t]
  \begin{center}
  \centerline{\includegraphics[width=3.5in]{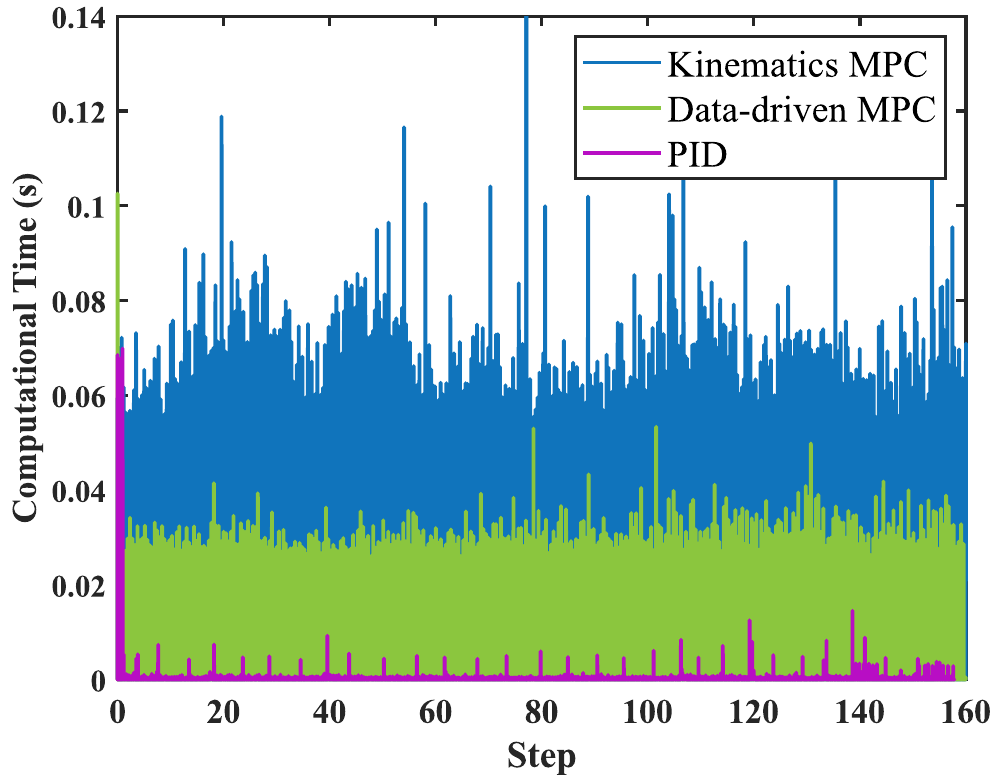}}
  \caption{Computational time of Data-driven MPC, PID and kinematics MPC.}
  \label{fig 6}
  \end{center}
\end{figure}

From Fig.~\ref{fig 6}, it can be observed that the computation time of DDMPC is almost half that of the vehicle kinematics MPC, since the computation time of it is only related to the data volume $N$ and the prediction horizon $L+v$, and not to the complexity of the system. This shows that we can ensure control accuracy while minimizing computation time by choosing appropriate values for $N$, $L$, and $v$.

\section{Conclusion}

In this study, we researched the existing Data-driven Model Predictive Control method proposed by \cite{coulson2019data, berberich2020data, berberich2020robust, berberich2020trajectory} and apply it to steering control of autonomous vehicle. Our experiments demonstrated that the algorithm could achieve stable front wheel angle control for tracking the reference trajectory, and compared to traditional MPC algorithms, it effectively reduces control errors and computation time. For more demonstrations of the effects of the experimental section go to: https://john0915aaa.github.io/DDMPC-for-AV-steering/.

Our future work will focus on enhancing the algorithm’s robustness and real-time adaptability to further improve its effectiveness in diverse driving conditions.


\bibliographystyle{unsrt}
\bibliography{ref}

\end{document}